\documentclass[10pt,journal,twocolumn]{IEEEtran}
\usepackage{times,epsfig,amsbsy,amssymb,float,color}
\usepackage[cmex10]{amsmath}
\usepackage{mdwmath, mathrsfs, bbm}
\usepackage{mdwtab}
\usepackage{amsfonts}
\usepackage{subfigure}
\usepackage{multicol}
\usepackage{amsmath,amssymb}
\usepackage{cases}
\usepackage{booktabs}
\usepackage{algorithm}
\usepackage{stfloats}
\usepackage{color}
\usepackage{algorithmicx}
\usepackage{algpseudocode}
\usepackage{mathtools}
\usepackage{authblk}
\usepackage{cite}
\usepackage[numbers,sort&compress]{natbib}

\usepackage{flushend}

\usepackage{stackengine}
\usepackage{scalerel}
\def\clipeq{\!\mathrm{=}\!}
\def\Lla{\Longleftarrow\!\!\!\!}
\def\Lra{\!\!\!\!\Longrightarrow}
\newcommand\xleftrightarrows[3][0]{%
  \mathrel{%
  \ifthenelse{\equal{#1}{0}}%
  {\stackunder[2pt]{\stackon[3pt]{$\Lla\Lra$}%
     {$\scriptstyle#2$}}{$\scriptstyle#3$}}%
  {\stackunder[2pt]{\stackon[3pt]{$\Lla\hstretch{#1}{\clipeq}\Lra$}%
     {$\scriptstyle#2$}}{$\scriptstyle#3$}}%
  }%
}

\input epsf
\title{An Adaptive Optimal Mapping Selection Algorithm for PNC using Variable QAM Modulation}
\author{Tong~Peng, Yi~Wang, Alister G. Burr and Mohammad Shikh-Bahaei \vspace{-2em}

\thanks{T. Peng was with the Department of Electronics, University of York, and now is with the Centre for Telecommunications Research, Department of Informatics, King's college London, UK (e-mail: tong.peng@york.ac.uk, tong.peng@kcl.ac.uk). Y. Wang and A. G. Burr are with the Department of Electronics, University of York, UK	
(e-mails: yi.wang@york.ac.uk, alister.burr@york.ac.uk). M. Shikh-Bahaei is with the Centre for Telecommunications Research, Department of Informatics, King's college London, UK (e-mail: m.sbahaei@kcl.ac.uk).

This research is funded by EPSRC NetCoM project EP/K040006/1 and partially by EPSRC IoSIRE project EP/P022723/1.}
}

\begin{document}
\maketitle\pagestyle{empty}
\begin{abstract}
Fifth generation (5G) wireless networks will need to serve much higher user densities than existing 4G networks, and will therefore require an enhanced radio access network (RAN) infrastructure. Physical layer network coding (PNC) has been shown to enable such high densities with much lower backhaul load than approaches such as Cloud-RAN and coordinated multipoint (CoMP). In this letter, we present an engineering applicable PNC scheme which allows different cooperating users to use different modulation schemes, according to the relative strength of their channels to a given access point. This is in contrast with compute-and-forward and previous PNC schemes which are designed for two-way relay channel. A two-stage search algorithm to identify the optimum PNC mappings for given channel state information and modulation is proposed in this letter. Numerical results show that the proposed scheme achieves low bit error rate with reduced backhaul load.
\end{abstract}


\section{Introduction}\label{sec:introduction}

The concept of network multiple input, multiple output (N-MIMO) \cite{M.V.Clark} has been known for some time as a means to overcome the inter-cell interference in fifth generation (5G) dense cellular networks, by allowing multiple access points (APs) to cooperate to serve multiple mobile terminals (MTs). This was implemented in the coordinated multipoint (CoMP) approach standardized in LTE-A \cite{D.Lee}. More recently the Cloud Radio Access Network (C-RAN) concept has been proposed, which has similar goals \cite{R.Irmer}. However these approaches result in large loads on the backhaul network (also referred to as fronthaul in C-RAN) between APs and the central processing unit (CPU), many times the total user data rate. 

While there has been previous work addressing backhaul load reduction in CoMP and C-RAN, using, for example, Wyner-Ziv compression \cite{re-bkload0}, iterative interference cancellation \cite{re-bkload2}, or compressive sensing \cite{nonidcran}, the resulting total backhaul load remains typically several times the total users' data rate. A novel approach was introduced in \cite{Lattice_QTSun, DongF}, based on physical layer network coding (PNC), which reduces the total backhaul load to a level in the range of the total users' data rate. However the research on PNC has mostly focused on two-way relay channel (TWRC) application \cite{Popovski, DongF2} or lattice code-based PNC design \cite{CF_Nazer}, which have significant disadvantages in terms of engineering applicability in N-MIMO networks. Moreover all these schemes have assumed that the same modulation order is used by all MTs sharing an AP, and that the signal strength at an AP is the same from all the MTs.


In this letter, we consider a more practical scenario in which the MTs may transmit with different powers. In this case, different modulation schemes are employed at each MT to maintain the required BER performance. In the simulations, a comparison of the proposed algorithm with the CoMP approaches is given, along with the impact of the estimated channel coefficients to the accuracy of optimal matrix selection. The primary contributions of this letter are summarized as follows: (1) an adaptive mapping selection algorithm when a different QAM modulation scheme is employed at each MT; (2) binary mapping matrices are employed so that the proposed algorithm can be implemented in current practical systems to achieve engineering applicability; (3) the dimension of mapping matrices is minimised to achieve lower computational complexity and higher degree of freedom.  


\section{System Model}
\begin{figure}[t]
  \centering
\begin{minipage}[t]{1\linewidth}
\centering
  \includegraphics[width=0.7\textwidth]{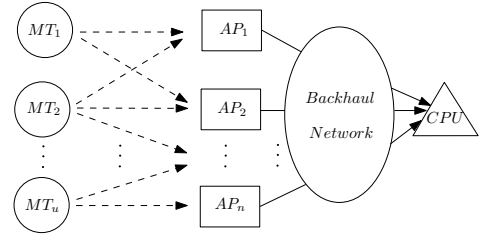}
\end{minipage}
\caption{\small The uplink system diagram.} \label{fig:system}\vspace{-1em}
\end{figure}

A two-stage uplink system model for N-MIMO is illustrated in Fig.\ref{fig:system}, where $u$ MTs are served by $n$ APs, and the APs are connected to a CPU via a backhaul network. A single antenna is provided at all MTs and APs. The first stage in the uplink is called the multi-access stage, in which all MTs broadcast symbols to APs at the same time. Each AP maps the superimposed signal to a PNC codeword vector and forwards it to the CPU via a backhaul network. The multi-access stage is assumed to use wireless communications, whilst the backhaul link is a lossless but capacity-limited `bit-pipe'. Note that the backhaul link can be wireless or fixed and the proposed scheme can be implemented in both cases.

A $2^m$-QAM modulation scheme is employed at each MT with a modulation function of $\mathscr{M}:\mathbb{F}_{2^m}\longrightarrow \Omega$, where $m$ stands for a modulation order, in bits per symbol, and $\Omega$ stands for the set of all possible modulated symbols. In the multi-access stage, $u_j$ MTs are served by an AP where the MTs transmit their signals simultaneously, yielding a received signal at the $j^{\mathrm{th}}$ AP of
\begin{align}\label{equ:sys_mod}
r_{j} = \sum_{i=1}^{u_j}h_{j,i}s_{i} + z_j, ~~ \mathrm{for} ~ j = 1,2,\cdots, n,
\end{align}
where $h_{j,i}$ denotes the complex Gaussian distributed random channel coefficient between the $j^{\mathrm{th}}$ AP and the $i^{\mathrm{th}}$ MT, and $s_i = \mathscr{M}(\mathbf{b}_i)$ is the $2^m$-QAM modulated signal, where $\mathbf{b}_i$ denotes the binary data vector with a dimension of $m \times 1$. $z_j$ denotes the additive white Gaussian noise with zero mean and variance $\sigma^2$.

For the purpose of industrial applicability, binary mapping matrices are employed in the proposed PNC encoder and decoder. Define a mapping function $\mathbf{G}_j$ at the $j^{\mathrm{th}}$ AP, the PNC encoding is then given by
\begin{align}\label{equ:pnc_encode}
\mathbf{x}_j = \mathbf{G}_j \otimes \mathbf{b},
\end{align}
where $\mathbf{b}\triangleq [\mathbf{b}_1, \mathbf{b}_2, \cdots, \mathbf{b}_{u_j}]^T$ denotes the $m_s \times 1$ binary joint message vector with $\mathbf{b}\in\mathbb{F}_2^{m_s \times 1}$ and $m_s = \sum_{i=1}^{u_j} m_i$, where $m_i$ denotes the modulation order at the $i^{\mathrm{th}}$ MT. $\mathbf{G}_j$ denotes a binary matrix with dimensions of $l_j \times m_s$, where $l_j$ denotes the number of rows of the mapping matrix used at the $j^{\mathrm{th}}$ AP, and $\otimes$ denotes matrix multiplication over $\mathbb{F}_2$. $\mathbf{x}_j \in\mathbb{F}_2^{l_j \times 1}$ is the network codeword vector (NCV) detected by the AP which consists of $l_j$ linear combinations of the original binary data.

We define $\mathbf{s}_{c_j}$ as a set which contains all possible combinations of the modulated signals from the MTs, where $u_j$ denotes the number of MTs served by the $j^\mathrm{th}$ AP. Given a $1 \times u_j$ channel coefficient vector $\mathbf{h}_j\triangleq [h_{j,1}, h_{j,2}, \cdots, h_{j,u_j}]$ at the $j^{\mathrm{th}}$ AP, the vector containing all $u_j^{m_s}$ possible superimposed signals can be calculated by
\begin{align}\label{equ:sup_sig_vet}
\mathbf{s}_{j,\mathrm{sc}} \triangleq [s^{(1)}_{j,\mathrm{sc}}, s^{(2)}_{j,\mathrm{sc}}, \cdots, s^{(u_j^{m_s})}_{j,\mathrm{sc}}] = \mathbf{h}_j \mathbf{s}_{c_j}, ~ \mathrm{for} ~ j = 1,2,\cdots, n.
\end{align}
According to the definition, the modulation function $\mathscr{M}(\cdot)$ is a one-to-one bijective mapping function which maps all possible combinations of binary data to the complex symbols $s \in \Omega$. In (\ref{equ:pnc_encode}), the joint message vector $\mathbf{b}$ can be mapped to an NCV $\mathbf{x}$ by a binary matrix $\mathbf{G}$. Then with the expression in (\ref{equ:sup_sig_vet}), we can always find a surjective PNC mapping function $\mathscr{N}_j$ that maps the superimposed constellation point $s^{(k)}_{j,\mathrm{sc}}$ to an NCV, mathematically given by
\begin{align}
\mathscr{N}_j(s^{(k)}_{j,\mathrm{sc}})=\mathbf{x}_j, ~\mathrm{for} ~k=1,2,\cdots,u_j^{m_s}.
\end{align}
At each AP, an estimator calculates the conditional probability of each possible NCV given the mapping function $\mathscr{N}_j$ and the channel coefficients. The estimator returns the log-likelihood ratio (LLR) of each bit of $\mathbf{x}_j$ which is then applied to a soft decision decoder. In the backhaul stage, the NCVs will be forwarded to the CPU from each AP, and then by concatenating them and multiplying by the inverse of the binary PNC mapping matrix, the original data from each MT will be recovered. Note that the global mapping matrix $\mathbf{G}\triangleq [\mathbf{G}_1 ~\mathbf{G}_2 ~\cdots ~\mathbf{G}_{u_j}]^T$ must be non-singular in order to unambiguously recover the source data \cite{TCOM}. Maximum likelihood (ML) detection can also be used for PNC decoding \cite{Popovski}.



\section{Design Criterion for Binary PNC}
\label{sec:PNC.Design}

Singular fading in the multi-access stage is a serious problem which affects detection performance at the CPU. We give a simple example here with $2$ MTs to illustrate this problem and the singular fading is defined as a situation in which different pairs of transmitted signals cannot be distinguished at the receiver, mathematically given by
\begin{align}\label{equ:sfs}
h_{j,1}s_1 + h_{j,2}s_2 = h_{j,1}s'_1 + h_{j,2}s'_2,
\end{align}
where $s_i$ and $s'_i$ stand for the QAM modulated signals at the $i^{\mathrm{th}}$ MT, and $s_i \neq s'_i$. The special channel coefficient vector $\mathbf{h}_\mathrm{sf}\triangleq [h_{j,1}, h_{j,2}]$ is defined as a singular fade state (SFS). Note that the solution of (\ref{equ:sfs}) is not unique, since there is more than one SFSs for each QAM scheme. (5) implies that superimposed constellation points corresponding to two different MT data combinations coincide, and hence this data cannot be unambiguously decoded at this AP. 


Besides the coincident symbols, there is a set of superimposed symbols that would map to the same NCV; this set is defined as a \textit{cluster} and denoted by $\mathbf{s}_\mathrm{cl}=[s_{j,\mathrm{sc}}^{(1)}, s_{j,\mathrm{sc}}^{(2)}, \cdots]$. We then define the minimum distance between different clusters as 
\begin{align}\label{equ:dmin}
d_{\mathrm{min}} = &\min_{\mathscr{N}_j(s_{j,\mathrm{sc}}^{(\tau_i)})\neq \mathscr{N}_j(s_{j,\mathrm{sc}}^{(\tau^{\prime}_k)}) }  |s_{j,\mathrm{sc}}^{(\tau_i)} - s_{j,\mathrm{sc}}^{(\tau^{\prime}_k)}|^2, \\
\forall s^{(\tau_i)}_{j,\mathrm{sc}} \in \mathbf{s}_\mathrm{cl}^{(\tau)}, &~\forall s^{(\tau^\prime_k)}_{j,\mathrm{sc}} \in \mathbf{s}_\mathrm{cl}^{(\tau^\prime)}, ~\text{for} ~i=1,2,..., ~k=1,2,... \notag.
\end{align}
In the coincident symbols cases $d_\mathrm{min} = 0$. The design criterion is to employ a mapping function that labels the constellation points within a clash to the same NCV with maximised $d_{\mathrm{min}}$ in order to achieve unambiguous decoding at the CPU. A detailed mathematical proof is derived in \cite{TCOM} and the design criterion holds in multiple-MT case. 

\section{Adaptive Mapping Function Selection Algorithm}
\label{sec:CombInfo}
In this section, we describe an adaptive PNC mapping matrix selection algorithm based on the design criterion introduced in the previous section. According to the criterion, the separate mapping matrices used at each AP should encode the constellation points within one cluster to the same NCV and additionally, the global mapping matrix $\mathbf{G}$ formed by the concatenation of the selected mapping matrices at each AP should be invertible for unambiguous decoding at the CPU. We define an $L \times u_j$ matrix $\mathbf{H}_\mathrm{sfs}=[\mathbf{h}_{\mathrm{sfs}_1}, \cdots, \mathbf{h}_{\mathrm{sfs}_L}]^T$ whose rows contain all special channel vectors that cause different SFSs for $u_j$ modulation scheme combinations. Like the work in \cite{DongF}, our proposed algorithm comprises two procedures, the first of which is an Off-line search and the second is an On-line search. The proposed Off-line and On-line algorithms are described in Algorithm 1.


\begin{algorithm}[htb]
\caption{Binary Matrices Selection (Off/On-line Search Algorithm)}
\label{Alg:MGMR}
\begin{algorithmic}[1]
\Statex
\Statex {\textbf{Off-line Search}}
\State {Define $\mathbf{b}_{s}=[\mathbf{b}_{s_1} \cdots \mathbf{b}_{s_{u}}]$ as a $ u^{m_s}\times m_s$ matrix containing all possible combinations of binary data}
\State {Define $\mathbf{G}_\mathrm{ofl}$ as a set that contains all $l \times m_s$ binary matrices for $l \in [1,m_s]$, where the number of matrices is $N_\mathrm{ofl}$}
\State {$\mathbf{s}_c = [M_1(\mathbf{b}_{s_1}), \cdots, ~M_2(\mathbf{b}_{s_u})]^T$} 
\For {$i=1:L$}  \Comment{each SFS}
\State {$\mathbf{s}_{i,\mathrm{sfs}} \triangleq [s^{(1)}_{i,\mathrm{sfs}},\cdots,s^{(u^{m_s})}_{i,\mathrm{sfs}}]=\mathbf{h}_{\mathrm{sfs}_i}\mathbf{s}_c $}
\State {Calculate distances between the superimposed signals: $~~~~~~d = |s^{(n_1)}_{i,\mathrm{sfs}} - s^{(n_2)}_{i,\mathrm{sfs}}|^2$, $\forall n_1, n_2, n_1 \neq n_2$} 
\State {Find $N_\mathrm{cl}$ different clashes $\mathbf{s}_\mathrm{cl}$ containing the $s_{i,\mathrm{sfs}}$ with $d = 0$ and the corresponding binary data combinations $\mathbf{b}_\mathrm{cl}$}
\For {$n_\mathrm{ofl}=1:N_\mathrm{ofl}$} 					\Comment{each binary matrix}
\For {$n_\mathrm{cl} = 1 : N_\mathrm{cl}$} \Comment{each clash}
\State {$\mathbf{x}^{(n_\mathrm{cl})}_\mathrm{cl} = \mathbf{G}^{(n_\mathrm{ofl})}_\mathrm{ofl} \otimes \mathbf{b}^{(n_\mathrm{cl})}_\mathrm{cl} $} \Comment{NCV calculation}
\EndFor
\If {all $\mathbf{x}^{(n_\mathrm{cl})}_\mathrm{cl}$ are the same}
\State {Store $\mathbf{G}^{(n_\mathrm{ofl})}_\mathrm{ofl}$ as a matrix candidate}
\EndIf
\EndFor
\EndFor
\Statex {\textbf{On-line Search}}
\State {Define $\mathbf{G}_\mathrm{onl}$ as the set that contains the $N_\mathrm{sel}$ selected mapping matrix candidates}
\For {$j = 1 : J$} \Comment{each AP}
\State {$\mathbf{s}_{j,\mathrm{sc}} = \mathbf{h}_j\mathbf{s}_c $} \Comment{All possible superimposed signals}
\For {$n_\mathrm{sel}=1:N_\mathrm{sel}$} \Comment{each selected matrix}
\State {$\mathbf{x}^{(n_\mathrm{sel})}_\mathrm{sel} = \mathbf{G}^{(n_\mathrm{sel})}_\mathrm{onl} \otimes \mathbf{b}_s$} \Comment{NCV calculation}
\State {Find the set of clusters from the positions of repeated columns in $\mathbf{x}^{(n_\mathrm{sel})}_\mathrm{sel}$, and among all pairs of clusters:}
\State { $d_{\mathrm{min}} = \min\limits_{\forall i,k, i \neq k} |s^{(\mathrm{cl}_i)}_{j,\mathrm{sc}} - s^{(\mathrm{cl}_k)}_{j,\mathrm{sc}}|^2, ~~~~~~~~~~~~~~~~~~$ where $\mathrm{cl}_i$ and $\mathrm{cl}_k$ denote the cluster indexes}
\EndFor
\EndFor
\State {The non-singular global mapping matrix selection: $~~~~~\mathbf{G}=[\mathbf{G}_1 \cdots \mathbf{G}_J]^T$ with maximum $d_{\mathrm{min}}$}

\end{algorithmic}
\end{algorithm}


The main differences between the proposed design and the work in \cite{DongF} is as follows. In \cite{DongF}, MTs employ the same modulation scheme and the channels have the same average path loss, and the dimension of the mapping matrices stored at each AP is equal to $m_s/2$ so that the sizes of the NCVs at each AP are the same. Our approach focuses on the situation in which the modulated signals from MTs are received at an AP with different powers, then in order to maintain a good performance in terms of error probability and throughput, different modulation schemes should be used. In addition, in order to achieve a higher degree of freedom, mapping matrices with different dimensions may be utilised. The non-singular global mapping matrix used for PNC decoding at the CPU is formed by concatenating the mapping matrices used at each AP with the maximum $d_\mathrm{min}$. When the channel qualities between the MTs and the $j^{\mathrm{th}}$ AP are poor, the value of $d_{\mathrm{min}}$ may be much smaller than that at the other APs, and this leads to performance degradation. Our solution is to employ $l_j \times m_s$ mapping matrices at the $j^{\mathrm{th}}$ AP with $l_j \in [1, m_s]$. An $l \times m_s$ matrix with $l > m_s/2$ at one AP and $l < m_s/2$ at the other may give a greater overall minimum inter-cluster distance $d_\mathrm{min}$ over both APs than choosing $l = m_s/2$ at both. 

Our proposed algorithm provides a solution to reduce backhaul load in N-MIMO networks with a high user density by employing PNC technique. However, the calculation of singular fading with multiple MTs in PNC is still an open question in binary systems. In order to apply the proposed algorithm in practical scenarios, OFDM \cite{tradeoff_PC} could be utilised by allocating a pair of MTs to the same frequency in MAC stage which divides the MAC stage with multiple MTs into multiple MAC stages with two MTs. In \cite{Acs}, we have tested the proposed algorithm with OFDM scheme using multiple USRP boards. The synchronisation issue of the proposed algorithm in OFDM systems is studied and interference from other MTs is treated as additional noise which leads to a limited performance degradation when synchronisation error occurs.
 
According to \cite{TCOM}, the number of SFSs increases exponentially with increasing modulation order index $m$. This phenomenon leads to an increase in the number of matrices returned by the Off-line search and makes the On-line search algorithm impractical. However, the $2^m$-QAM modulation schemes are all defined in $\mathbb{F}_{2^m}$ and the same constellation points can be found in different constellation books. This property of $2^m$-QAM modulation schemes allows us to use the same mapping matrix to resolve different SFSs, and the number of mapping matrices being stored reduces dramatically. We term SFSs which can be resolved by the same matrix as another SFS, \textit{image SFSs}. For example, the original number of SFSs in QPSK is $13$ and that in $16$-QAM is $389$. After the image SFSs are removed, the number is reduced to $5$ in QPSK and $169$ in $16$-QAM. A detailed study of the impact to the network performance with less number of SFSs is derived in \cite{TCOM} with computational complexity analysis.

\section{Numerical Results and Discussion}
\label{sec:NumRes}

\begin{figure}[t]
  \centering
\begin{minipage}[t]{1\linewidth}
  \includegraphics[width=0.9\textwidth]{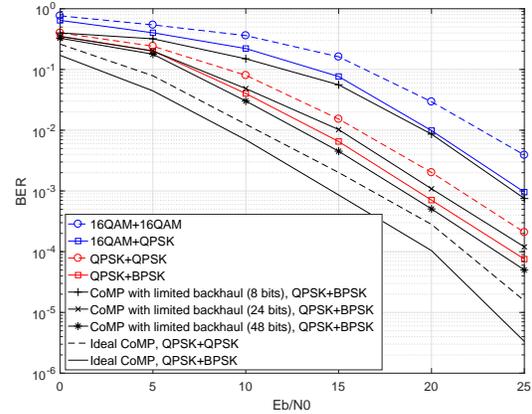} 
\end{minipage}
\caption{\small BER of the proposed algorithm with $3$dB difference in channel quality.} \label{fig:ber_BQ_16Q} 
\end{figure}

In this section, we illustrate the simulation results of the proposed algorithm in the $5$-node system shown in Fig. \ref{fig:system}. We assume the multi-access links are wireless and the backhaul is a capacity-limited wired link. For simulation simplicity, two MTs are served by two APs and we assume a $3$dB difference in average path loss between each AP and the two MTs. The MT with better channel quality employs a higher order modulation scheme, corresponding to a higher rate. A convolutional code is used in the simulation and other channel codes, such as LDPC \cite{Burr.MC}, may also be employed. As benchmarks we use ideal CoMP with unrestricted number of bits transmitted in backhaul network and non-ideal CoMP in which the soft information on the backhaul is quantized to a total of $12$, $6$ and $2$ bits per symbol, resulting in the total backhaul load of $48$, $24$ and $8$ respectively (since four LLRs are calculated at each AP).


In Fig. \ref{fig:ber_BQ_16Q}, the BER performances of different approaches are illustrated. We assume block fading in the multi-access stage and that perfect channel information can be obtained to obtain accurate SFSs. As shown in the figure, the proposed approach employs different modulation schemes achieves $3$dB and $2$dB improvement in BER performance compared to that with the same modulation schemes and CoMP, respectively. The ideal CoMP is the best among the approaches due to the availability of unrestricted bits in backhaul network; while quantization with limited precision results in a performance degradation in non-ideal CoMP. In the $QPSK+BPSK$ case, the proposed algorithm outperforms CoMP with $24$ bits, while requiring a total of only $9$ backhaul bits. A $2.5$dB improvement can be observed with $QPSK + BPSK$ compared to $QPSK + QPSK$ in this case; while $16QAM + QPSK$ gives an improvement of $3$dB compared to $16QAM + 16QAM$.


\begin{figure}[t]
  \centering
\begin{minipage}[t]{1\linewidth}
  \includegraphics[width=0.9\textwidth]{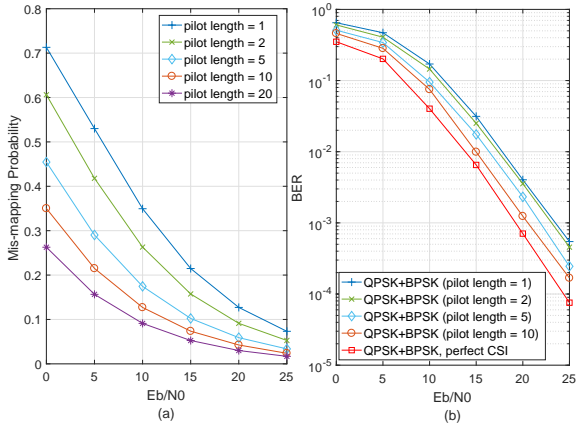} 
\end{minipage}
\caption{\small Non-optimal mapping matrices selection porbability with estimated channel.} \label{fig:cha_map_prob} 
\end{figure}

The channel coefficients obtained at each AP are important to the proposed algorithm because its performance relies on the accurate identification of the closest SFS. In practical communication networks, the trade-off between the length of information symbols and that of pilot symbols for channel estimation has been discussed in \cite{tradeoff_PC}. We have illustrated the effect of channel estimation on the proposed algorithm in terms of the probability of selecting a non-optimal matrix, which is shown in Fig. \ref{fig:cha_map_prob}. In the simulation, different numbers of pilot symbols have been employed to estimate the channel, which affects the accuracy of identification of the SFS calculated at each AP. As illustrated in (a), when a short pilot sequence is employed, there is a maximum of $70\%$ probability that the optimal mapping matrix will not be selected in a low SNR scenario which corresponds to a $5$dB degradation compared to that when perfect channel knowledge is employed. With a pilot sequence length of $5$, and at SNRs sufficient to ensure a low BER, the mis-mapping probability drops below $10\%$ and the performance degradation reduces to $3$dB. According to Fig. \ref{fig:cha_map_prob}, a pilot sequence with a length between $5 - 10$ is enough for the proposed adaptive selection algorithm. 

\section{Future Work and Conclusion}
In this letter, an engineering applicable approach is presented to implement adaptive PNC in binary N-MIMO systems when different MTs transmit with different powers. Unlike our work in \cite{TCOM} and \cite{Acs}, we focus on applying the proposed algorithm in practical scenarios in which multiple MTs could employ different modulation schemes. We also illustrate how the estimated channel affects the optimal matrix selection accuracy. The simulation results illustrate the benefits of the proposed PNC approach in terms of backhaul load reduction and BER improvement compared to the non-ideal CoMP with quantized bits in the backhaul channel. 

An extension study of applying the proposed PNC design in binary systems with full-duplex (FD) APs \cite{FD-D2D}-\cite{FD-ARQ} has been started. Practical PNC design with cross layer optimisation in \cite{CLD-D2D}-\cite{CLD-D2D_2} provide another research direction, and the research in \cite{SE} focuses on spectrum efficiency solutions and can be extended to PNC-implemented systems. Moreover, the PNC application with optimal resource allocation \cite{RA-CDMA} and \cite{MQAM} is critical in order to serve the 5G systems and achieve massive data transmissions with high accuracy and low latency.

\small      

\bibliographystyle{IEEEtran}
 {\footnotesize{}}

\end{document}